\documentclass[aps,prl,twocolumn,superscriptaddress]{revtex4}
\usepackage{mathtools}
\usepackage{graphicx}
\usepackage{lmodern}
\usepackage{xfrac}
\usepackage{hyperref}

\newcommand{\FWM}{\mathrm{FWM}}
\newcommand{\dete}{\mathrm{D}}

\newcommand{\tautheory}{\Delta t}

\begin{document}
\title{Upper bound on the duration of quantum jumps}

\author{Mathias~A.~Seidler}
\affiliation{Centre for Quantum Technologies, National University of Singapore, 3 Science Drive 2, Singapore 117543}

\author{Ricardo~Guti\'errez-J\'auregui}
\affiliation{Institute for Quantum Science and Engineering, Texas A\&M University, College Station, TX 77843, USA}

\author{Alessandro~Cer\`{e}}
\affiliation{Centre for Quantum Technologies, National University of Singapore, 3 Science Drive 2, Singapore 117543}

\author{Roc\'io~J\'auregui}
\email[]{rocio@fisica.unam.mx}
\affiliation{Instituto de F\'isica, Universidad Nacional Aut\'onoma de M\'exico and Apartado Postal 20-364, 01000, M\'exico D.F., M\'exico}

\author{Christian~Kurtsiefer}
\email[]{christian.kurtsiefer@gmail.com}
\affiliation{Centre for Quantum Technologies, National University of Singapore, 3 Science Drive 2, Singapore 117543}
\affiliation{Department of Physics, National University of Singapore, 2
  Science Drive 3, Singapore 117551}

\date{\today}
\begin{abstract}
We present a method to estimate the time scale of quantum jumps from the
time correlation of photon pairs generated from a cascade decay in an atomic
system, and realize it experimentally in a cold cloud of~$^{87}$Rb.
Taking into account the photodetector response, we find an upper bound for the duration of a quantum jump of~$21\pm11$\,ps.

\end{abstract}


\maketitle

\textit{Introduction --}
The concept of quantum jumps is traced back to Bohr and the old quantum
theory~\cite{Bohr:1913gl}. This theory found its bases on Planck's hypothesis of energy quanta and lead to successful explanations of the photoelectric effect, discrete atomic spectra, and distribution of a blackbody in thermal equilibrium. While it described the discrete states of systems accurately, the theory raised questions regarding the transition periods---or quantum jumps---between states. The periods occurred at random times and, by considering intermediate states forbidden, they had to be instantaneous. The general inadequacy of the old quantum theory appeared to be fixed by wave mechanics~\cite{Schrodinger_1952}; yet, for quantum systems under observation, it has not been possible to depart completely from the quantum jump concept~\cite{Bell}.

The idea for early observations of quantum jumps was seeded on Dehmelt's electron shelving proposal~\cite{Dehmelt:1975}, where the fluorescence of a driven two-state system is abruptly interrupted as the system transitions to a third, metastable, state.
This scheme has been implemented for single trapped ions~\cite{Nagourney:1986ih,Sauter:1986dx,Bergquist:1986zz} and neutral atoms~\cite{Finn:1986dx}. Recent quantum jump experiments involve nuclear spins~\cite{Neumann_2010}, cavity--\cite{Gleyzes:2007jr} and circuit~\cite{Vijay:2011gl,Zlatko_2018} quantum electrodynamics architectures.
In all these demonstrations, the  evolution of the monitored variable has the form of a telegraphic signal~\cite{Cook:1985ft} presenting on and off times whose duration can be described quantitatively through a study of the waiting time-distribution~\cite{CohenTannoudji:1986dt} and applying photodetection theory~\cite{Glauber:1963,Javanainen_1986}.
Since the probability for a system to be found in a particular state is conditioned to a given measurement record, the inferred  transition times  reflect the way the system is monitored~\cite{Carmichael, Wiseman_1993}.
Apart from studies based on the photoelectric effect~\cite{Ossiander:2016fx,Ossiander:2018is},
the time structure of the rise and descend transitions has been rarely explored at the limits of contemporary experimental capabilities~\cite{Zlatko_2018}.

In this work
we focus on
quantum jumps associated to spontaneous transitions involving discrete states.
We consider an alternative configuration for the observation of quantum jumps in atomic systems: a monitored cascade three-level system~\cite{Javanainen_1986,Hendriks:1988ev}.
Contrary to the shelving configuration, where the transition to the metastable
state is inferred from the absence of a fluorescent signal, i.e., a null
measurement~\cite{Pegg:1991hp,Porrati:1987}, state information in a cascade system is acquired through the detection of correlated photon pairs with a well-defined time ordering.
This, coupled with the fact that phase matching conditions allow for photon
coincidences to be accurately measured, makes the cascade configuration ideal
to experimentally investigate the discontinuous changes in  atomic states.

We observe the quantum jumps in the second order time correlation function of
the fluorescence generated by a cascade-based four-wave mixing (FWM) in a cold
ensemble of~$^{87}$Rb, and
model the experiment as a cascade three-level system via the adiabatic elimination of one of the intermediate states.
To improve the timing precision for observing any possible jump dynamics, we
take into account the measured impulse response of the single photon detectors
for estimating an upper bound on the duration of the quantum jump from the measured time correlation.

\begin{figure}
 \centering
\includegraphics[width=1\columnwidth]{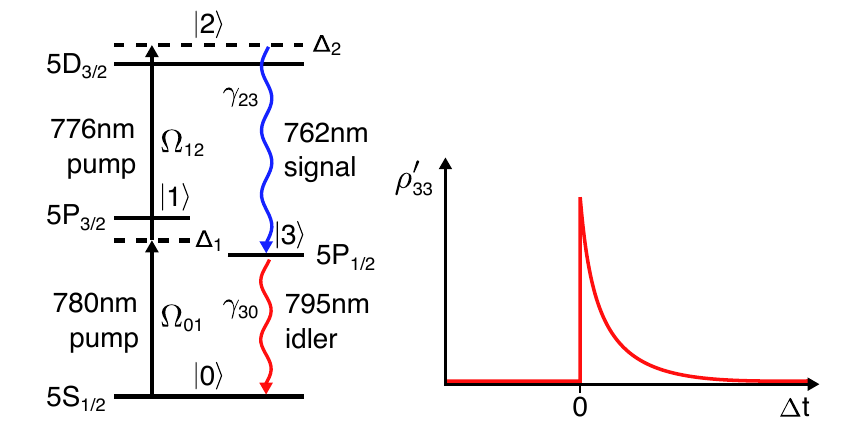}
 \caption{\label{fig:diamond}
(Left)
Atomic level configuration in a four-wave mixing experiment.
(Right)
Time evolution of the population of level $\vert 3\rangle$, conditioned on the detection of a signal photon according to Eq.~(\ref{eq:qj}).
}
\end{figure}

\textit{Theory --}
Consider a four-level system in the diamond configuration as depicted in Fig.~\ref{fig:diamond}. For~$^{87}$Rb, the atomic ground state $5S_{1/2}$,
level $\vert 0\rangle$, is coupled to an excited state $5D_{3/2}$, $\vert 2\rangle$, through a two-step excitation and decay process involving the intermediate levels $5P_{3/2}$, $\vert 1\rangle$, and $5P_{1/2}$, $\vert 3\rangle$. The excitation path $\vert 0\rangle \leftrightarrow \vert 1\rangle \leftrightarrow \vert 2\rangle$ is coherently driven by a pair of non-resonant lasers, while coupling to the electromagnetic environment induces the de-excitation path $\vert 2\rangle \rightarrow \vert 3\rangle \rightarrow \vert 0\rangle$ through spontaneous emission of idler and signal photons.
Due to the collective nature of the process, a phase-matching condition of the
cascade emission allows to efficiently collect the light used to monitor the
atomic state.

The joint probability distribution
to detect a heralding photon in mode $s$ at time $t$
and a correlated photon in mode $i$ at time $t + \tautheory$ is
related to the correlation function
\begin{equation}
 C(t, t + \tautheory) =  \langle \hat{a}^{\dagger}_{s}(t)\hat{a}^{\dagger}_{i}(t+ \tautheory) \hat{a}_{i}(t + \tautheory)\hat{a}_{s}(t) \rangle\, \, ,
\end{equation}
with the annihilation and creation operators  $\hat{a}_{s,i}$ and
$\hat{a}^{\dagger}_{s,i}$ of the $s$ and $i$ modes.
For an electromagnetic environment in the vacuum state,
the correlation function~$C$ is proportional to
the atomic polarization correlation
\begin{equation}
 {P}(t,t+\tautheory) = \langle \hat{\sigma}^\dagger_{23}(t) \hat{\sigma}^\dagger_{30} (t+\tautheory) \hat{\sigma}_{30} (t+\tautheory) \hat{\sigma}_{23}(t)\rangle\,,
\end{equation}
with the lowering operator $\hat{\sigma}_{ij} \equiv \vert j \rangle \langle i \vert $ describing transitions from the atomic level $\vert i\rangle$ to level $\vert j\rangle$ in the Heisenberg picture. Using the quantum regression theorem~\cite{Lax:1968ig} it is possible to show that
\begin{equation}\label{intro:eq_1}
 {P}(t,t+\tautheory) = \rho_{22}(t)\rho^\prime_{33}(t+\tautheory) \, ,
\end{equation}
so that atomic density matrix element $\rho_{22}(t)$ represents the population of state $\vert 2\rangle$ at time $t$, while $\rho^{\prime}_{33}(t+\tautheory)$ that of
state  $\vert 3\rangle$ at time $t+\tautheory$ under the initial condition $\rho^\prime_{33}(t)=\vert 3\rangle\langle 3\vert$. This result reflects the well-defined time order that allows the study of quantum jumps in the four-wave mixing process.

It is possible to obtain an analytical expression for ${P}(t,t+
\tautheory)$; its derivation, however, is greatly simplified if the
intermediate level $\vert 1 \rangle$ is adiabatically eliminated. This
elimination is valid for far-detuned lasers, which induce rapid oscillations
on the probability amplitude to find the system in the intermediate level and
allow us to consider its zero average value. Under this approximation, the
system maps to an effective three-level cascade system with an evolution ruled by the master equation:
\begin{equation}
 \dot{\rho} = (i \hbar)^{-1} \left[ \mathcal{H}_{\text{eff}}, \rho \right] + \sum_{n,m} \gamma_{nm} \left[ 2 \hat{\sigma}_{nm} \rho\hat{\sigma}_{nm}^{\dagger} - \hat{\sigma}_{nn} \rho - \rho \hat{\sigma}_{nn} \right]  \,,\label{eq:bloch}
\end{equation}
with the effective Hamiltonian
\begin{equation}
 \mathcal{H}_{\text{eff}} = \hbar \Delta_{\text{eff}} \vert 2 \rangle \langle 2 \vert + \sum_{i=2,3}E_i \vert i \rangle \langle i \vert + \frac{\hbar \Omega_{\text{eff}}}{2} \left( \vert 2 \rangle \langle 0 \vert + \vert 0 \rangle \langle 2 \vert \right)\,,
\end{equation}
and
effective detuning and Rabi frequency
\begin{equation}
 \Delta_{\text{eff}}  = \Delta_{2} + \frac{\Omega_{01}^{2}}{4 \Delta_{1}} -  \frac{\Omega_{12}^{2}}{4 \Delta_{1}}\, \quad\text{and}\quad
 \Omega_{\text{eff}}  = - \frac{\Omega_{01}\Omega_{12}}{2 \Delta_{1}}\, .
\end{equation}
In these expressions, $\Omega_{ij}$ denote the bare Rabi frequencies for the
induced $\vert i\rangle \leftrightarrow \vert j \rangle$ transitions,
$\gamma_{ij}$ the spontaneous decay rates, and~$\Delta_{i}$ the detuning
between the driving lasers and the atomic resonances~$(E_i-E_j)/\hbar$.
The master equation evolution is equivalent to a solvable set of algebraic equations obtained using the Laplace transform,
allowing
for the calculation of analytical expressions that describe the evolution
of the density matrix. With the initial condition  $\rho^\prime_{33}(t_0) = 1$
where $t_0$ is the time at which the signal photon is detected, and for the
experimental parameters shown below one finds
\begin{equation}
\label{eq:qj}
\begin{aligned}
 \rho^\prime_{33}(t_0 + \tautheory) & = 0  \quad   & {\mathrm {for}}\, & \tautheory<0\,, \\
 \rho^\prime_{33}(t_0 + \tautheory) & \approx
 e^{-\left(\gamma_{30} + \tfrac{\gamma_{23}}{2}\right) \tautheory}
   \quad  & {\mathrm {for}}\, & \tautheory>0\,,
\end{aligned}
\end{equation}
while~$\rho_{22}(t)$ is a continuous function of $t$. Thus,  the correlation function  $P(t,t+\tautheory)$ exhibits a discontinuity at~$\tautheory = 0$ that reflects
the breaking of time symmetry induced by the quantum jump associated to the transition $\vert 2\rangle \rightarrow \vert 3\rangle$.
For $\tautheory>0$ the atom can perform a second jump to the ground state
with statistics given by~$\rho^\prime_{33}$ in Eq.~(\ref{eq:qj}).

Signal and idler photons impinge on photon detectors that, while able to
distinguish between either mode, are unable to determine with certainty the
hyperfine level they were emitted from. This indeterminacy causes that the
atomic state conditioned to the detection of a signal photon is given by a
superposition of  the hyperfine states of level $\vert 1\rangle$, i.e.,
$5P_{1/2}$ $F=1$ and $F=2$ states. This yields  oscillations of
$\rho^\prime_{33}(t_0 + \tautheory)$ for~$\tautheory>0$ with a frequency
determined by the beating frequencies of these hyperfine
levels~\cite{Gulati:2015ee}. These quantum beats can be well reproduced with our
model by numerically solving the Bloch equations for a density matrix that involves all hyperfine sublevels of the $\vert i \rangle$,
i=0,1,2,3 states. The frequency and amplitude of the quantum beats
are highly sensitive  to the polarization, intensity and detuning of the pump
lasers. Details of the calculations will be provided in a forthcoming
publication. Important to notice here is that the observation of quantum beats provides a striking evidence of quantum coherence.

\begin{figure}
 \centering
 \includegraphics{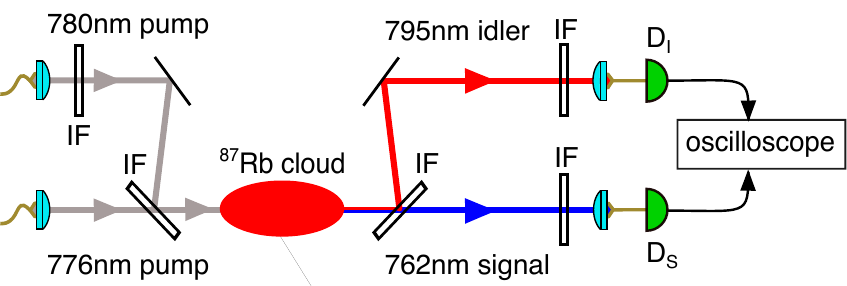}
 \caption{\label{fig:schematic}
  Schematic of the four-wave mixing experiment. IF: interference filters to combine pump beams  and to separate the photons pairs;
  \mbox{D$_S$, D$_I$}: silicon avalanche photodiodes (APD).}
\end{figure}

\textit{Experiment --}
Figure~\ref{fig:schematic} shows schematically the experimental setup for
generating time-ordered photon pairs by four-wave mixing in a cold ensemble
of~$^{87}$Rb atoms.
Pump beams at 780\,nm and 776\,nm excite atoms from the $5S_{1/2},F=2$
ground level to the $5D_{3/2},F=3$ level via a two-photon transition. The
signal (762\,nm) and idler (795\,nm) photons emerge from a cascade decay
back to the ground level through the $5P_{1/2}$ level, and are coupled to single mode fibers.
Phase matching between pump and target modes is ensured with
all four modes propagating in the same direction.
The two pump modes are almost collimated and have a Gaussian beam waist of
about 400\,$\mu$m in the atomic cloud. The linearly polarized pump mode at
780\,nm is red-detuned by 40\,MHz from the $5S_{1/2},F=2$ to $5P_{3/2},F=3$ transition and has an optical power of 0.5\,mW.
The orthogonally polarized pump mode at 776\,nm shares the same optical mode,
has an optical power of 6.5\,mW, and is tuned such that the two-photon
excitation is blue-detuned by 4\,MHz from the difference between the ground
state and the $5D_{3/2}, F=3$ level.
We record the detection event time differences of the photon pairs with a
digital oscilloscope (sampling rate $4\times10^{10}\,\text{s}^{-1}$) with an
effective time resolution
below 10\,ps; the single photon avalanche detectors themselves have a nominal
timing jitter around 50\,ps FWHM.
\begin{figure}
 \includegraphics[width=1\columnwidth]{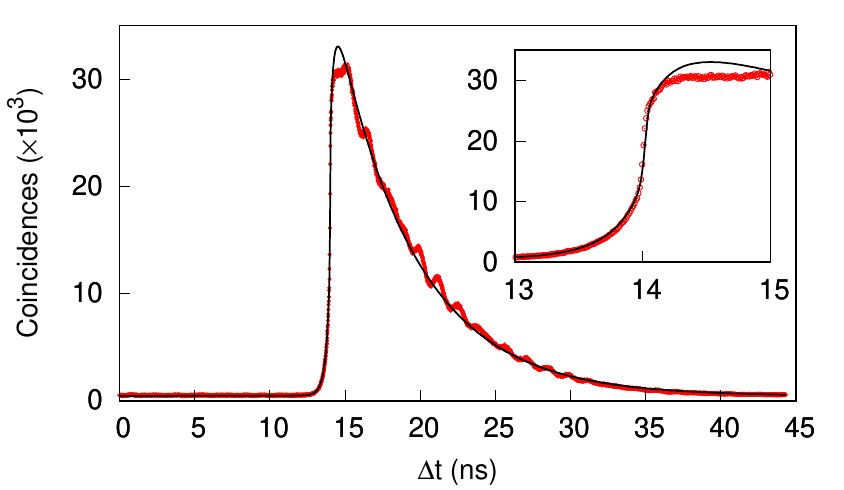}
 \caption{\label{fig:FWM_G2}
   Histogram \(G_{\FWM}(\Delta t)\) of detection time differences for photons
   pairs generated by four-wave mixing in the cold cloud of~$^{87}$Rb.
   The continuous line shows the result of the best fit of
   Eq.~(\ref{eq:fit_model}). Inset: detail of the sharp rise corresponding to a quantum jump.}
\end{figure}

Figure~\ref{fig:FWM_G2} shows the histogram~$G_{\FWM}(\Delta t)$ of signal
and idler photodetection time differences~$\Delta t$ into 10\,ps wide bins.
The expected exponential decay starts at $\Delta t_0\approx14$\,ns due to
technical delays through fibers and cables. The decay time constant is shorter
than the natural decay time $1/\gamma_{30}\approx 27$\,ns and determined by the
number of atoms involved and the detuning of the pump fields~\cite{Srivathsan:2013,Cere:2018tt}.
The oscillations on top of the exponential decay are due to quantum beats between the transition of interest and alternative decay paths involving nearby hyperfine levels~\cite{Gulati:2015ee}.
The onset of the fluorescence in the idler mode, reflecting the quantum jump,
shows a rise time much faster than the time scales of the transitions
involved, commensurate with the time jitter of the avalanche photodetectors
(see inset of Fig.~\ref{fig:FWM_G2}).

To improve the time resolution for observing the quantum jump, we characterize the detector response function.
\begin{figure}
\includegraphics[width=1\columnwidth]{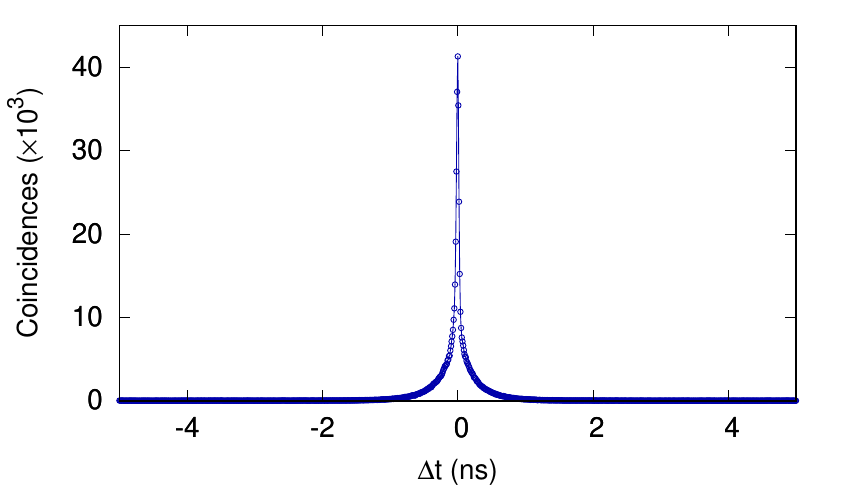}
 \caption{\label{fig:det_response}
  Histogram \(G_{\dete}(\Delta t)\) of the difference in detection time for
  photons pairs generated by spontaneous parametric down-conversion
  in a nonlinear optical crystal.
  The
  bandwidth of the photon pair exceeds 20\,nm, corresponding to a coherence
  time of about 0.1\,ps.
  Thus, we can take the measured time correlation as a reasonable
  approximation for the impulse response of the two-detector system.
 }
\end{figure}
For this, we direct photon pairs with a large optical bandwidth generated by
spontaneous parametric down-conversion (SPDC) in a nonlinear optical crystal
onto the two avalanche photodetectors. The SPDC source
is based on $\beta$-Barium Borate, cut for non-degenerate type-I
phase matching. Pumped with light at 405\,nm, it generates time-correlated
photon pairs at 770\,nm and 854\,nm~\cite{Villar:18}.
The measured bandwidths (23\,nm and 32\,nm, respectively) correspond
to a coherence time of the down-converted photon pairs below
0.1\,ps,
a negligible contribution to ~$G_{\dete}(\Delta t)$ in comparison to the detector response time scale.
The resulting calibration coincidence histogram $G_{\dete}(\Delta t)$ into
10\,ps wide bins
is shown in
Fig.~\ref{fig:det_response}.
Despite the wavelength difference, $G_{\dete}(\Delta t)$ does
not show any appreciable asymmetry. From this, we infer that the timing
response of the detectors does not vary significantly over this wavelength
range, and we thus expect that the measured behavior is a good approximation of the detector timing characteristics for photons at 762\,nm and 795\,nm.

\textit{Result --}
While the structure of~$G_{\FWM}(\Delta t)$  away from~$\Delta t = 0$ is well understood,
we have no model for a possible dynamic of the jump itself.
In order to associate a time scale to the transient behavior, we join
the two parts of
Eq.~(\ref{eq:qj}) with a
smooth heuristic transition
\begin{equation}\label{eq:sigmoid}
    \sigma (x)\:=\:\frac{1}{1+\exp\left(-x\right)}\,.
\end{equation}
The choice of~$\sigma (x)$ is not inspired by a specific dynamical model, its only purpose is to establish a time scale for the transient. This particular form is attractive since it admits the step function as a limiting scenario.
The atomic state, with the exponential decay with time constant~$\tau$ described by Eq.~(\ref{eq:qj}), is then enclosed in the monitor function
\begin{equation}\label{eq:jump_model}
    y(\Delta t; \alpha, \tau) \:=\:
    \sigma\!\left(\frac{\Delta t}{\alpha}\right)
    \,\exp\left( -\frac{\Delta t}{\tau}\right)\,,
\end{equation}
where
the jump timescale is characterized by~$\alpha$.

By convolving the monitoring function $y(\Delta t)$ in
Eq.~(\ref{eq:jump_model}) with
the normalized measured detector
response~$g_{\dete}(\Delta t) = G_{\dete}(\Delta t) / \sum G_{\dete}(\Delta
t)$,
we construct a model~$Y$ for the {\em
observed}~$G_{\FWM}(\Delta t)$ in Fig.~\ref{fig:FWM_G2},
\begin{equation}\label{eq:fit_model}
    Y(\Delta t)\:=\: A\:y(\Delta t - \Delta t_0; \alpha, \tau)*g_{\dete}(\Delta t) + Y_0\,
\end{equation}
with amplitude~$A$,
decay time~$\tau$,
accidental coincidence background~$Y_0$,
time delay~$\Delta t_0$, and characteristic time~$\alpha$ for the jump.
We then use $A$, $Y_0$, $\Delta t_0$, and $\alpha$ as free parameters to fit
$Y(\Delta t)$ to our measured time difference distribution.
This fit, shown as continuous line in Fig.~\ref{fig:FWM_G2},
results in $\alpha = 4.7 \pm 2.5$\,ps corresponding to
a~10\%--90\% rise time associated with the jump of~$21\pm11$\,ps.

\textit{Conclusion --}
We establish a bound for the duration of a quantum jump based on
the observed onset of time correlations between photons emitted from an atomic
cascade decay
in a cold cloud of~$^{87}$Rb. We find a value that is about three orders of
magnitude shorter than the natural lifetime of the involved atomic states, and
four orders of magnitude longer than an optical cycle.

In comparison with other techniques~\cite{Nagourney:1986ih,Sauter:1986dx,Bergquist:1986zz,Finn:1986dx,Neumann_2010,Gleyzes:2007jr,Vijay:2011gl,Zlatko_2018},
there seems to be no fundamental limit to the time resolution of this method
down to the time scale of the photoelectric effect.
We believe our measurement is still limited by the uncertainty in the time
response of the avalanche photodetectors, and potentially far from the
timescale of quantum jumps -- should there be one. Adoption of faster and
better characterized detectors,
for example from a recent generation of superconducting
nanowires~\cite{Shcheslavskiy:2016cr}, has the potential to significantly
improve the time resolution of such an experiment, and possibly establish or
abandon a resolvable time scale for quantum jumps.

\section*{Acknowledgments}
We like to thank A.~Villar and A. Ling for lending their photon pair source for
detector calibration. R.~J. acknowledges support from Conacyt LN-293471, A.~C.
acknowledges travel support from C{\'a}tedra {\'A}ngel Dacal. This work was supported by the Ministry of Education in
Singapore and the National Research Foundation, Prime Minister's office
(partly under grant no NRF-CRP12-2013-03).

\bibliographystyle{apsrev4-1}
%

\end{document}